\documentclass[final,5p,times,twocolumn]{elsarticle}

\usepackage{subfigure}
\usepackage{lineno,hyperref}
\modulolinenumbers[5]

\usepackage{amstext}
\usepackage{siunitx}

\DeclareSIUnit{\neq}{n_{eq} }
\DeclareSIUnit{\C}{c}
\DeclareSIUnit{\protons}{protons}

\journal{NIM-A}

\begin{document}

\begin{frontmatter}

\title{Results on Proton-Irradiated 3D Pixel Sensors Interconnected to RD53A Readout ASIC}

\author[ifca,cern]{J. Duarte-Campderros\corref{thecorrespondingauthor}}\ead{jorge.duarte.campderros@cern.ch}
\address[ifca]{Instituto de Fisica de Cantabria (Universidad de Cantabria/CSIC), Av. de los Castros s/n 39005 Santander, Spain}
\address[cern]{European Organization for Nuclear Research (CERN), 1211 Geneva 23, Switzerland}
\cortext[thecorrespondingauthor]{Corresponding author. Tel.: +41 22 76 71657}
\author{on behalf of the CMS Collaboration}
\author[ifca,cern]{\\E. Curr\'as}
\author[ifca,cern]{M. Fern\'andez,}
\author[ifca]{G. G\'omez}
\author[ifca]{A. Garc\'ia}
\author[ifca]{J. Gonz\'alez}
\author[ifca,cern]{E. Silva}
\author[ifca]{I. Vila}
\author[ifca]{R. Jaramillo}
\author[firenze]{M. Meschini}
\author[firenze]{R. Ceccarelli}
\author[milano,milano-uni]{M. Dinardo}
\author[milano,milano-uni]{S. Gennai}
\author[milano,milano-uni]{L. Moroni}
\author[milano,milano-uni]{D. Zuolo}
\author[torino]{L. Demaria}
\author[torino]{E. Monteil}
\author[pavia,bergamo]{L. Gaioni}
\author[pisa]{A. Messineo}
\author[trento,trento-uni]{G. F. Dalla Betta}
\author[trento,trento-uni]{R. Mendicino}
\author[fbk]{M. Boscardin}
\author[cnm]{S. Hidalgo}
\author[cnm]{A. Merlos}
\author[cnm]{G. Pellegrini}
\author[cnm]{D. Quirion}
\author[cnm]{M. Manna}
\address[firenze]{INFN Sezione de Firenze, Via Giovanni Sansone, 1, 50019 Sesto Fiorentino FI, Italy}
\address[milano]{INFN Sezione di Milano-Bicocca, Milano, Italy}
\address[milano-uni]{University of Milano Bicocca, Milano, Italy}
\address[torino]{INFN Sezione di Torino, Via Pietro Giuria, 1, 10125 Torino, Italy}
\address[pavia]{INFN Sezione di Pavia, Via Agostino Bassi, 6,  27100 Pavia, Italy }
\address[bergamo]{Universita degli studi di Bergamo, Bergamo, Italy }
\address[pisa]{INFN Sezione di Pisa and Universita di Pisa, Largo Bruno Fibonacci, 2, 56127, Pisa, Italy}
\address[trento-uni]{Universit\`a degli Studi di Trento, Via Calepina, 14, 38122 Trento, Italy}
\address[trento]{INFN Sezione di Padova, Gruppo Collegato di Trento, Trento, Italy}
\address[fkb]{FBK-Fondazione Bruno Kessler, Via Sommarive, 18 - Povo 38123 Trento, Italy}
\address[cnm]{Centro Nacional de Microelectr\'onica, IMB-CNM (CSIC), 08193 Barcelona, Spain}

\begin{abstract}
Test beam results obtained with 3D pixel sensors bump-bonded to the 
RD53A prototype readout ASIC are reported. Sensors from FBK (Italy) and IMB-CNM (Spain)
have been tested before and after proton-irradiation to an equivalent 
fluence of about \SI{1e16}{\neq\centi\metre\tothe{-2}} (\SI{1}{\mega\electronvolt} equivalent neutrons). This is
the first time that one single collecting electrode fine pitch 3D sensors are irradiated
up to such fluence bump-bonded to a fine pitch ASIC. The preliminary analysis of the collected
data shows no degradation on the hit detection efficiencies of the tested sensors
after high energy proton irradiation, demonstrating the excellent radiation tolerance 
of the 3D pixel sensors. Thus, they will be excellent candidates for the extreme radiation environment
at the innermost layers of the HL-LHC experiments.
\end{abstract}


\end{frontmatter}


\section{Radiation Hardness and 3D Columnar Pixel Sensors}
\label{sec:HL-LHC}
Pixel detectors in the innermost layers of the High Luminosity Large Hadron 
Collider (HL-LHC) experiments will have to survive a fluence in excess
of \SI{1e16}{\neq\centi\metre\tothe{-2}} (\SI{1}{\mega\electronvolt} equivalent neutrons),
while preserving high tracking efficiency~\cite{tdr}. It has already been 
demonstrated~\cite{prev_rad} that 3D pixel sensors are sufficiently radiation tolerant 
to resist the expected fluence on the innermost layers of the 
tracking systems of HL-LHC experiments, although it was done by using a coarse pitch 
readout chip. 
This work, complementing ref.~\cite{marco}, uses for the first time 
3D sensors bump-bonded to a fine pitch readout ASIC (RD53A). In addition,
the study presents characterization results of 3D pixel sensors of 
\SI{25 x 100}{\micro\metre\square} 
pitch with one single collecting electrode per cell, irradiated 
up to \SI{1E16}{\neq\cm\tothe{-2}}. 

\begin{figure}[!htbp]
\centering
\includegraphics[width=0.7\linewidth]{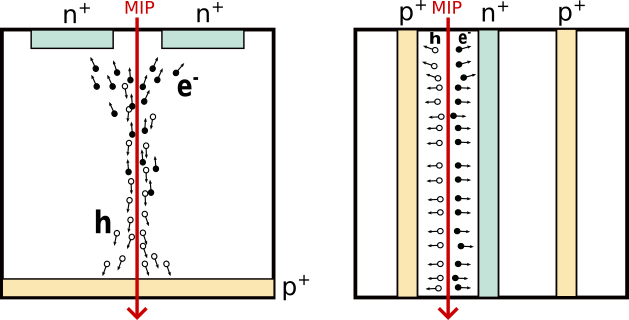}
\caption{Schematic diagram of a planar (left) and a 3D (right) sensor, showing
the charge carrier creation when a particle passes through, and the carriers movement
before being collected on the electrodes.}
\label{fig:planar_vs_3d}
\end{figure}

Columnar 3D pixel sensors~\cite{3d,3d_cnm} possess several 
intrinsic properties making them good candidates to create the layers of a tracking
system exposed to an extreme radiation environment. For example, the voltage
needed to fully deplete the sensor is much lower than in an equivalent planar sensor,
particularly after irradiation, due to small distance between junction and ohmic 
contacts. For the same reason, charge carriers travel less distance than in
an equivalent planar detector, therefore they are less impacted by trapping. Charge carriers 
in the 3D pixel sensors travel only \SI{35}{\micro\metre} for a \SI{50 x 50}{\micro\meter\square} 
pixel pitch independently of the sensor thickness, which is the driving parameter for planar pixels.
The 3D pixels have also a very fast response time.
Figure~\ref{fig:planar_vs_3d} compares both planar and 3D designs. 

\begin{figure}[htbp]
    \centering
    \subfigure[Single-sided 3D FBK] 
    {
        \includegraphics[width=0.45\linewidth,height=2.5cm]{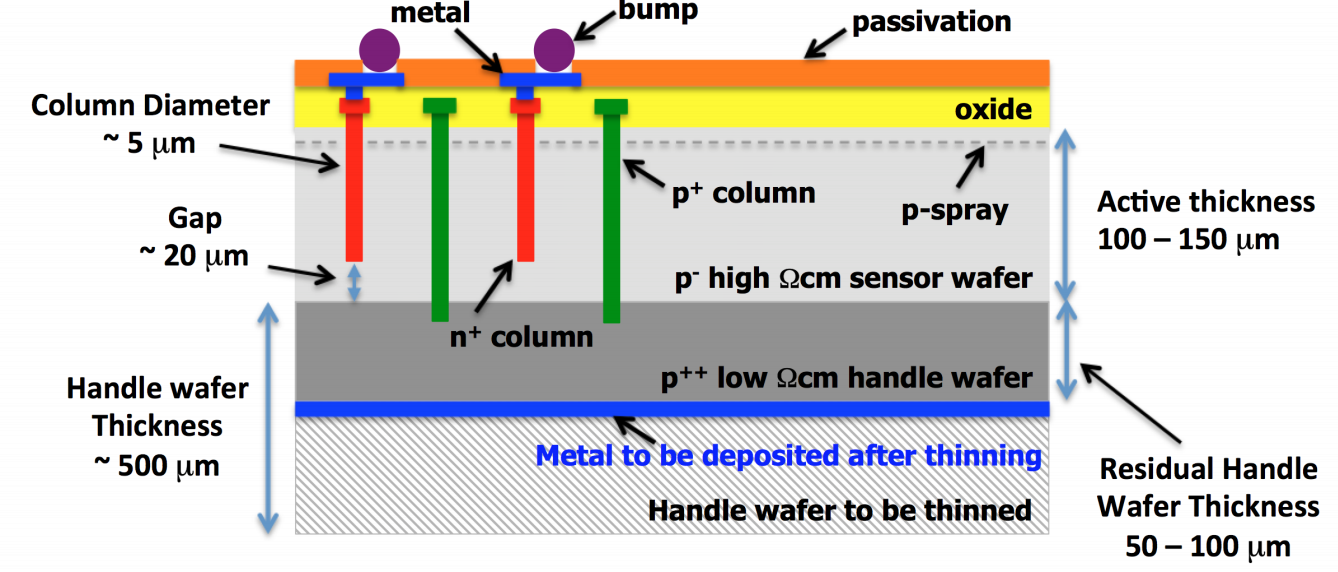}
    }
    \subfigure[Double-sided 3D CNM] 
    {
        \includegraphics[width=0.45\linewidth]{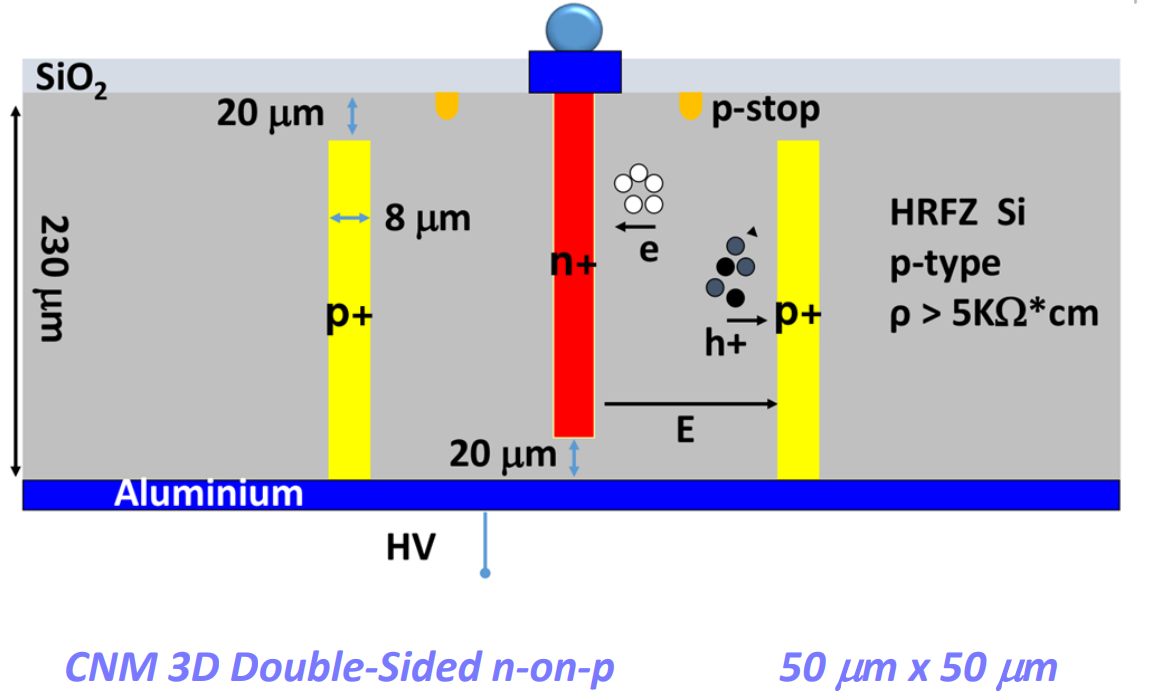}
    }
    \caption{Cross-section of the 3D pixel sensor layouts}
    \label{fig:sketches}
\end{figure}

The 3D sensors used in this work were fabricated at the FBK in Trento (Italy),
and in the IMB-CNM in Barcelona (Spain). Sensors from FBK were bump-bonded to 
the RD53A~\cite{rd53} readout-chip, while sensors from IMB-CNM were bump-bonded to the
ROC4SENS~\cite{roc4sens} chip. The substrates were p-type Si-Si 
high-resistivity wafers with \SI{130}{\micro\metre} thickness in the case of 
the FBK sensors, and conventional p-type high-resistivity FZ wafers of \SI{230}{\micro\metre}
thickness, for the IMB-CNM sensors. Detector and fabrication description can be found on 
ref.~\cite{3d,3d_cnm}. Figure~\ref{fig:sketches} shows both single-sided FBK and 
double-sided IMB-CNM pixel layout sensors.

\begin{figure}[htbp]
    \centering
    \subfigure[$25\times100\,\mu{\rm m}^2$ 2E sensor] 
    { 
            \includegraphics[width=0.45\linewidth,height=4.5cm,keepaspectratio]{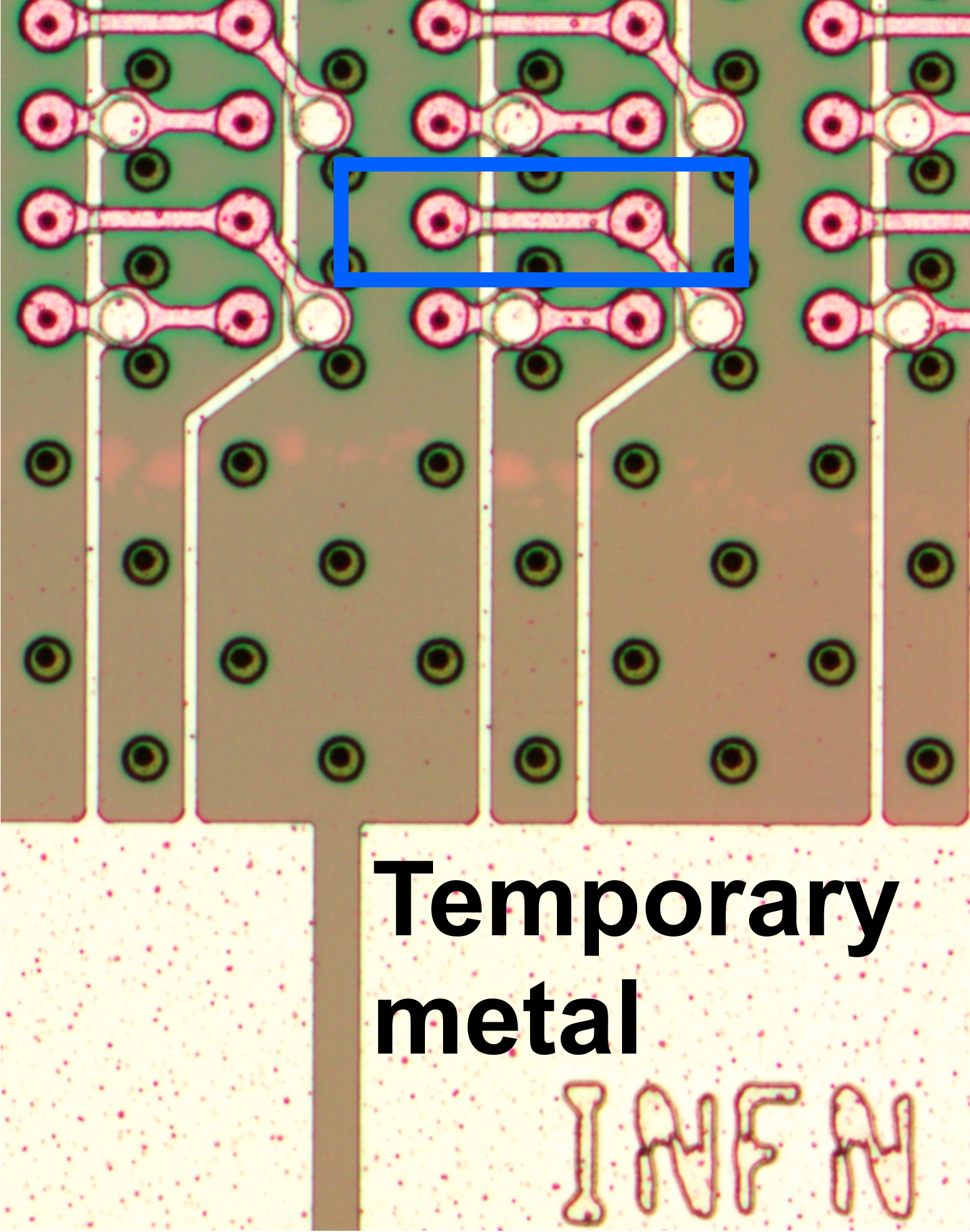}
    }
    \hspace{4mm}
    \subfigure[$50\times50\,\mu{\rm m}^2$ sensor]
    {
        \includegraphics[width=0.45\linewidth,height=4.5cm,keepaspectratio]{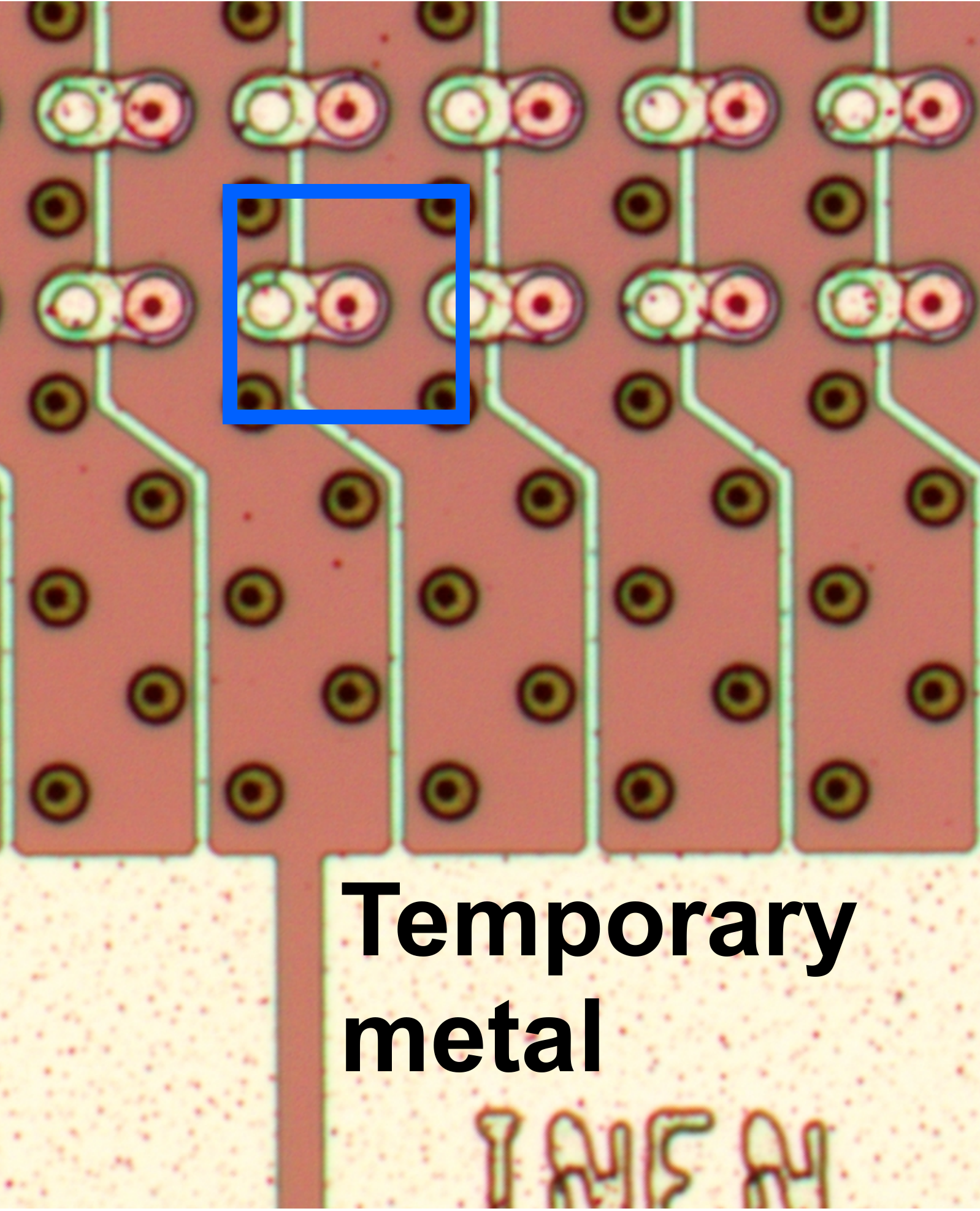}
    }
    \caption{Microscope pictures of 3D FBK sensors. The contact pad for the probe and
    the vertical metal lines connecting all pixels are visible in the pictures by the 
    \emph{Temporary metal} label, used for sensor testing at FBK premises and which is subsequently
    removed. A 3D single pixel cell on each sensor is highlighted with the blue frames. 
    Note the metal routing configuration in (a) in order to connect the readout electrodes to 
    the square matrix RD53A bumping pads.}
    \label{fig:pixphoto}
\end{figure}

The sensors were created with two different pixel cells: \SI{50 x 50}{\micro\metre\square} 
and \SI{25 x 100}{\micro\metre\square}, the latter having one (1E) or two (2E) collecting electrodes
per cell. The 1E case is particularly interesting given the large yield when producing them.
Two examples of pixel cells are shown in figure \ref{fig:pixphoto}. 
Both cell sizes are presently under evaluation in the Compact Muon Solenoid (CMS) collaboration
for the inner layers of the pixel detectors for HL-LHC. 

After fabrication, the FBK pixel sensor wafers were processed
for UBM (Under Bump Metallization), thinned down to \SI{200}{\micro\metre} 
total thickness, diced and bump-bonded to RD53A prototype chips at IZM (Berlin, Germany). 
The RD53A chip has 76800 readout channels (400 rows and 192 columns with a bump 
pad pitch of \SI{50 x 50}{\micro\metre\square}) and measures \SI{20.0x11.8}{\milli\meter\square}.
The pixel sensor bonded to the readout chip needs to be glued 
and wire-bonded onto an adapter card in order to be tested; these units will be
referred to as modules in the following text. All results referring to the
FBK+RD53A modules were obtained with the Linear Front-End~\cite{fe} (FE) in the central zone of RD53A 
(136 columns wide, from 128 to 263).

IMB-CNM sensors were also processed for UBM and then bump-bonded to ROC4SENS readout chips.
The ROC4SENS is a generic chip able to readout without zero suppression, and
therefore specially well suited for sensor studies. The chip has a bump pad pitch of 
\SI{50 x 50}{\micro\meter\square} with 160 rows and 155 columns providing 24800 pixels in
a \SI{9.8 x 7.8}{\milli\metre\square} surface. 

\section{Irradiation and Test Beam setup}
\label{sec:Irr}
Irradiations were performed in 2017 for the IMB-CNM sensors, and in 2018 for the FBK sensors 
at the CERN IRRAD facility~\cite{irrad}, a high intensity 
\SI[per-mode=symbol]{24}{\giga\electronvolt\per\C} proton beam which has a 
FWHM of \SI{12}{\milli\metre} in x and y directions. The target fluence for the IMB-CNM
sensors was \SI{3e15}{\neq\centi\metre\tothe{-2}}, and \SI{1e16}{\neq\centi\metre\tothe{-2}}
for the FBK ones. The latter modules were tilted on the IRRAD
beam at an angle of \SI{55}{\degree} in order to homogeneously irradiate the 
\SI{20 x 12}{\milli\metre\square} sensors and readout chip areas. 
The corresponding total ionizing dose was \SI{6}{\mega\gray} for \SI{1.65e16}{\protons\centi\metre\tothe{-2}}.

\begin{figure} [htbp]
    \centering
    \subfigure[Modules mounted on the tilted supports]
    {
        \includegraphics[width=0.45\linewidth,height=4.5cm,keepaspectratio]{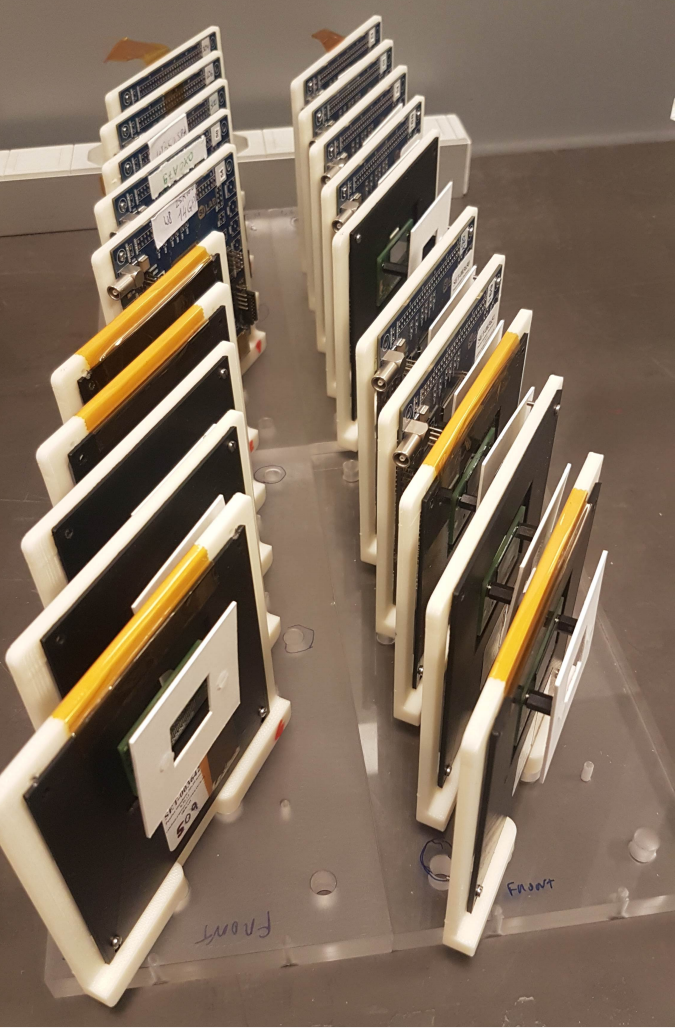}
    }
    \hspace{4mm}
    \subfigure[A module after irradiation]
    {
        \includegraphics[width=0.45\linewidth,height=4.5cm]{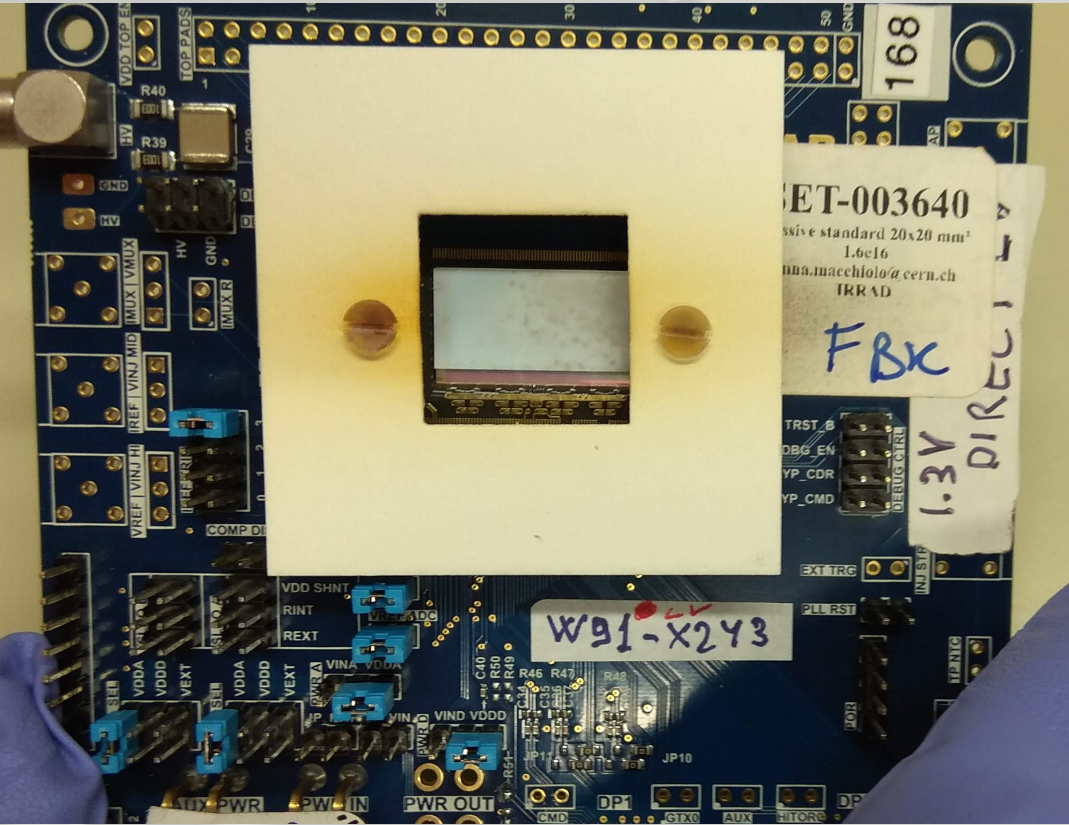}
    }
    \caption{Tilted modules mounted on the irradiation tray (a). A  module after
    irradiation (b), the dark brown band on the cardboard frame and on the nylon 
    screw heads is due to the proton beam passing through the tilted
    module.}\label{fig:irrad}
\end{figure}

Visual inspections after irradiation and data analysis are showing that the FBK modules
were displaced with respect to the irradiation beam axis by a few millimeters
and therefore the nominal fluence was reached only in a part of the modules, in particular
in about half of the linear FE. Several measurements and cross-checks are being 
performed in order to establish the effective fluence integrated on the modules.
All results shown here are based on the nominal requested equivalent fluence. 
Figure~\ref{fig:irrad} shows several modules mounted on the supports ready to be irradiated
and one of the modules characterized in this work.

RD53A modules were tested in two test beam experiments at the CERN North Area H6B 
before and after irradiation in July and October 2018 respectively. The test beam 
facility at CERN provided for this study \SI[per-mode=symbol]{120}{\giga\electronvolt\per\C} hadrons, 
allowing to characterize the sensor response to minimum ionizing particles. Particle trajectories are measured by
the high-spatial-resolution planes of an EUDET-type telescope~\cite{eudet_tel} with 
less than \SI{5}{\micro\metre} accuracy, and spatially correlated with the hits measured at our 
modules, i.e. Devices Under Test (DUT). An extra plane is placed between the telescope planes
and is used to correlate in time the telescope with the DUT hits. The hit efficiency
of a DUT is calculated from the available trajectories with a hit in the reference
sensor, and looking for a hit in the DUT. More details on an equivalent test beam setup
and data analysis can be found at ref.~\cite{equiv_tb}.

Irradiated modules were kept cold at temperatures between  $-20{\rm ^{\circ}C}$ and
$-30{\rm ^{\circ}C}$ using dry ice bricks. The temperature was monitored via PT1000
sensors located close to the backside of the module and via NTC resistors soldered 
on the adapter card. Both sensors gave consistent measurements. 

\begin{figure}[!htbp]
    \centering
    \subfigure[Module Threshold]
    {
        \includegraphics[width=0.45\linewidth,height=5cm]{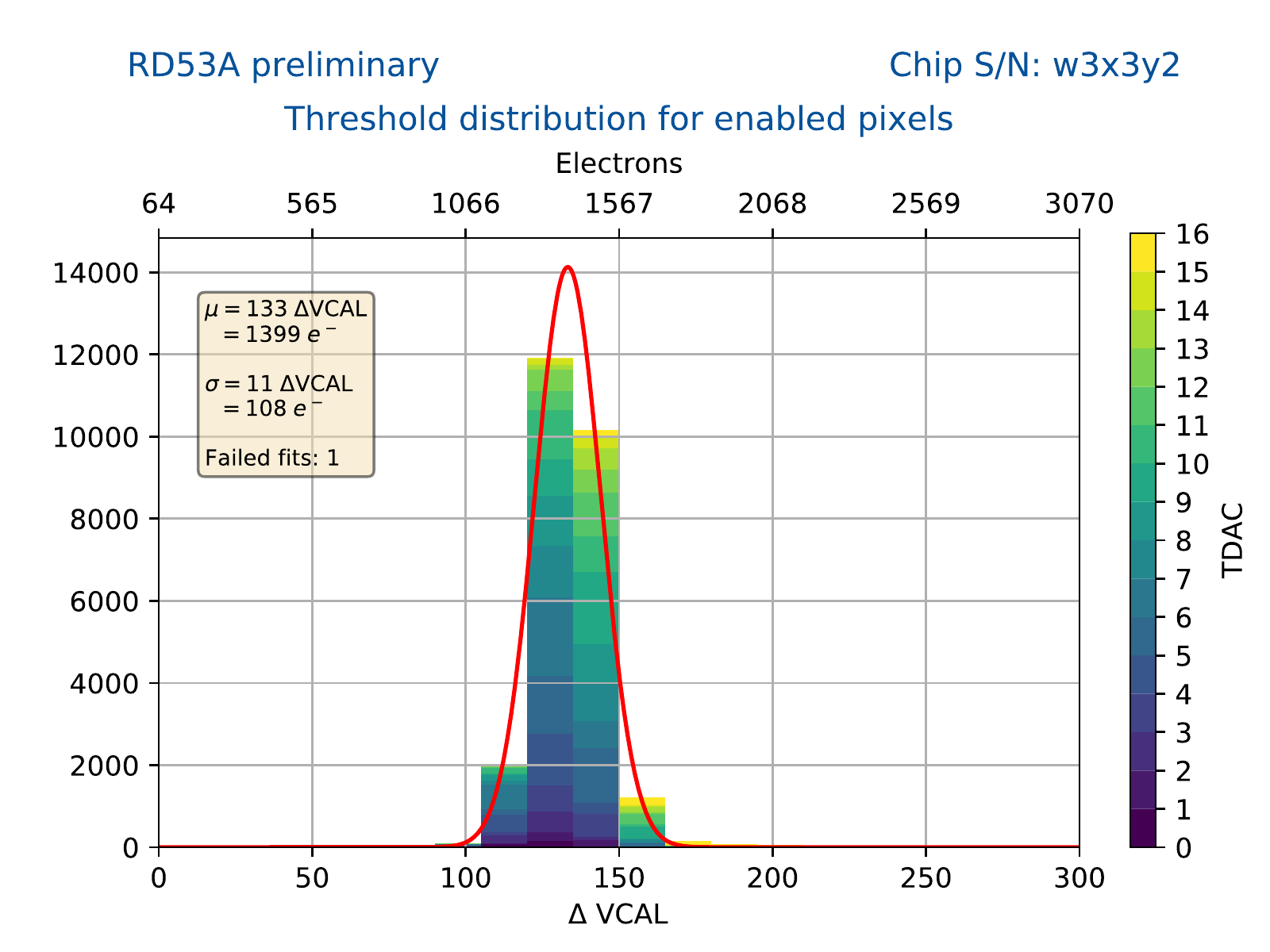}
    }
    \subfigure[Module noise]
    {
        \includegraphics[width=0.45\linewidth,height=5cm]{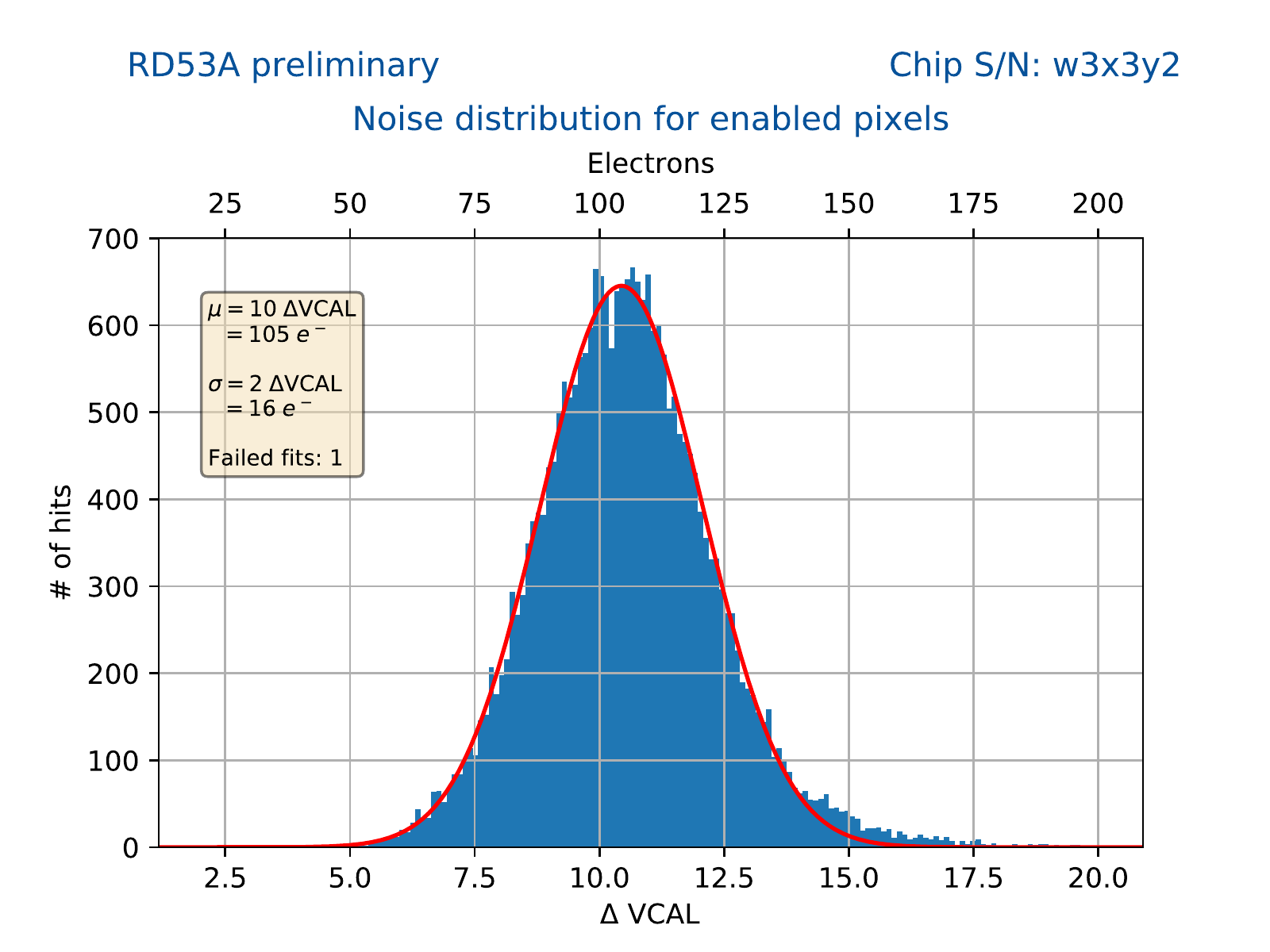}
    }
    \caption{Signal threshold (a) and noise (b) distributions for a 
    \SI{25 x 100}{\micro\metre\square} module after irradiation.}
    \label{fig:tunning}
\end{figure}

The readout chip parameters were tuned in order to reach low thresholds and 
noise, having at most 1.5\% masked pixels because of noisy channels. For the 
irradiated modules the average signal threshold was set  to about 1400 electrons, 
with a noise value of 105 electrons  for non-masked pixels, as shown in 
figure~\ref{fig:tunning} for a \SI{25 x 100}{\micro\metre\square}. Similar parameters
were found for the \SI{50 x 50}{\micro\metre\square}.
The color scale for the threshold distribution represents the 4-bit DAC
value used for the trimming of each individual pixel response.

\section{Sensor response and Results}
The 3D modules prior to irradiation reached hit detection efficiencies above \SI{98.5}{\percent}
for perpendicular incident tracks, already at moderate bias (less than \SI{15}{\volt}). 

\begin{figure}[!htbp] 
    \centering
    \subfigure[Efficiency before irradiation at \SI{3}{\volt} bias]
    {
        \includegraphics[width=0.90\linewidth,height=0.2\linewidth]{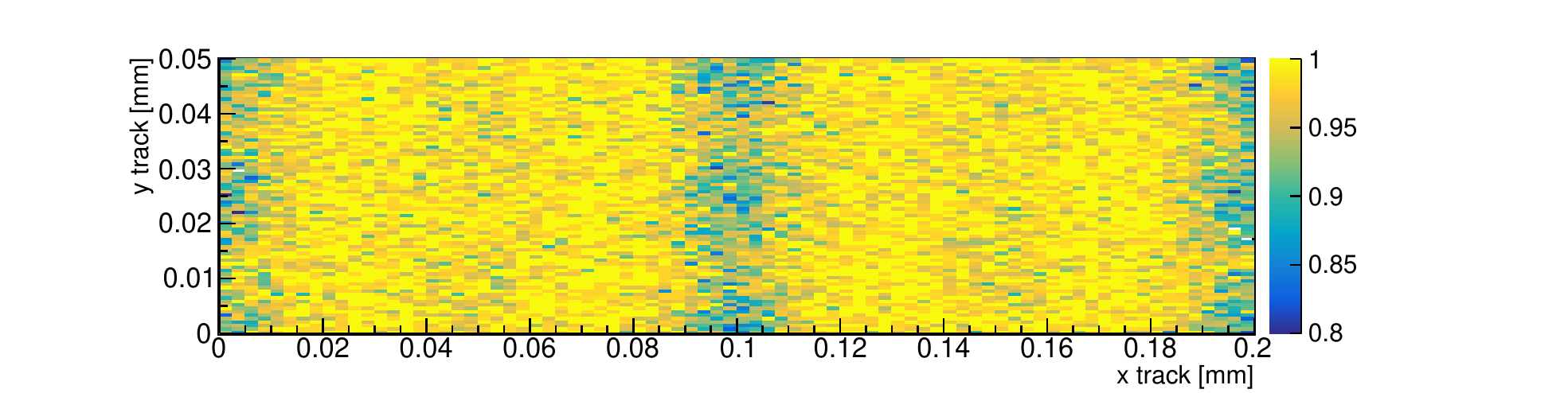}
    }
    \subfigure[Efficiency after irradiation at \SI{120}{\volt} bias]
    {
        \includegraphics[width=0.90\linewidth,height=0.2\linewidth]{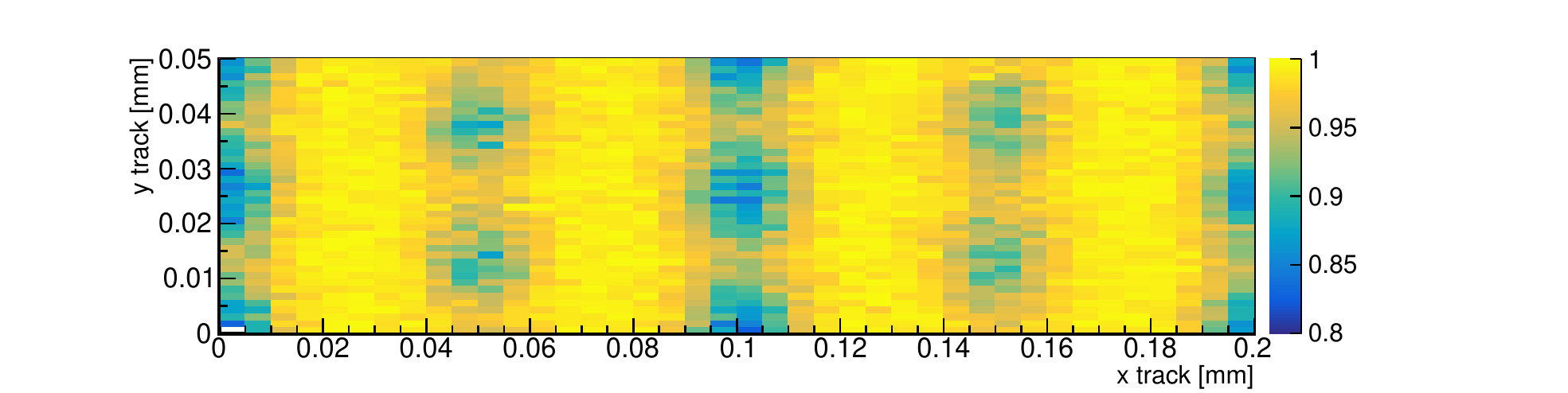}
    }
    \caption{Hit detection efficiencies before (a) and after irradiation (b) for a
    \SI{25 x 100}{\micro\metre\square} module.}
    \label{fig:s25x100}
\end{figure}

\begin{figure}[!htbp] 
    \centering
    \subfigure[Efficiency before irradiation at \SI{15}{\volt} bias]
    {
        \includegraphics[width=0.45\linewidth,height=0.45\linewidth]{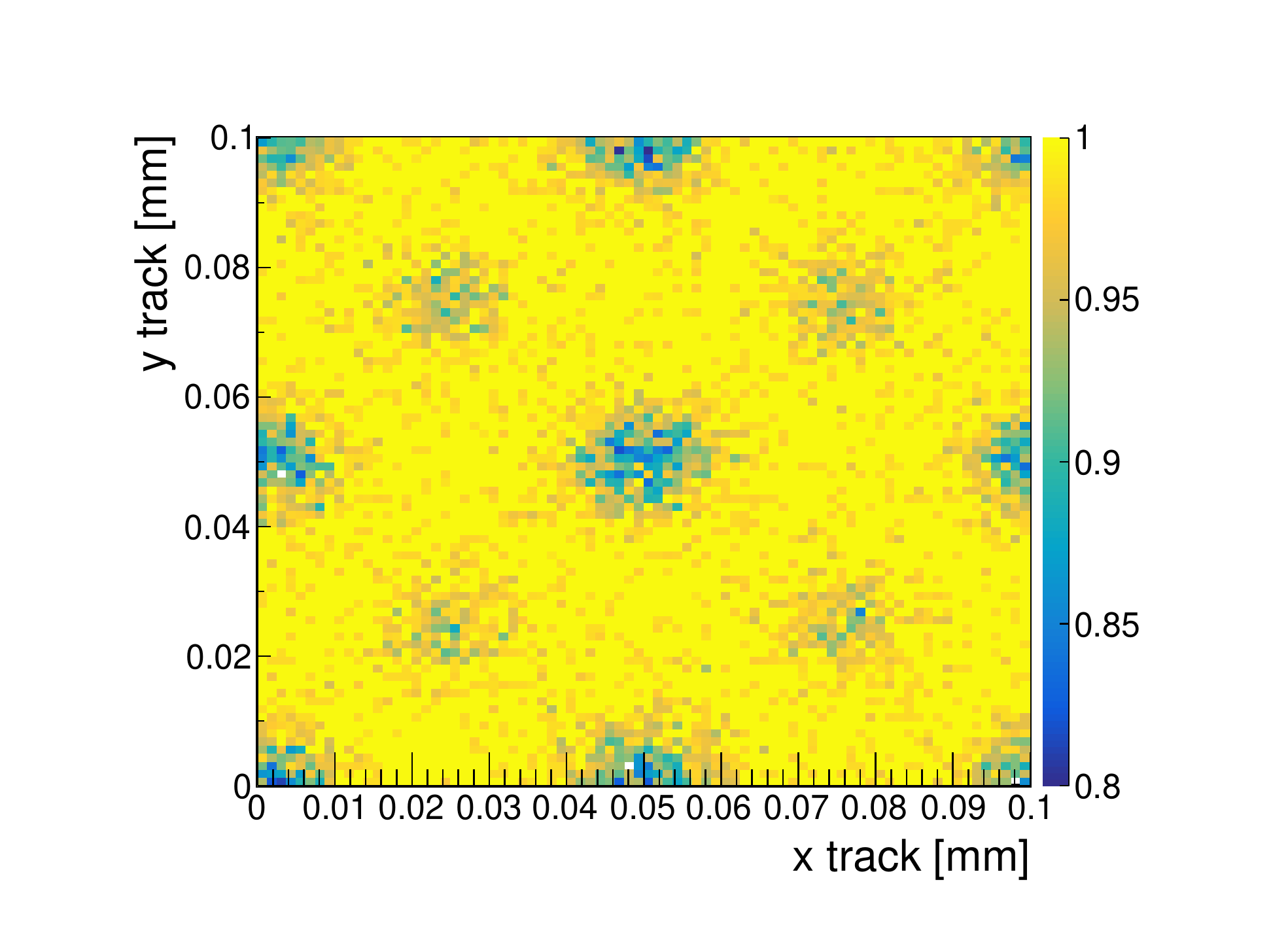}
    }
    \subfigure[Efficiency after irradiation at \SI{150}{\volt} bias]
    {
        \includegraphics[width=0.45\linewidth,height=0.45\linewidth]{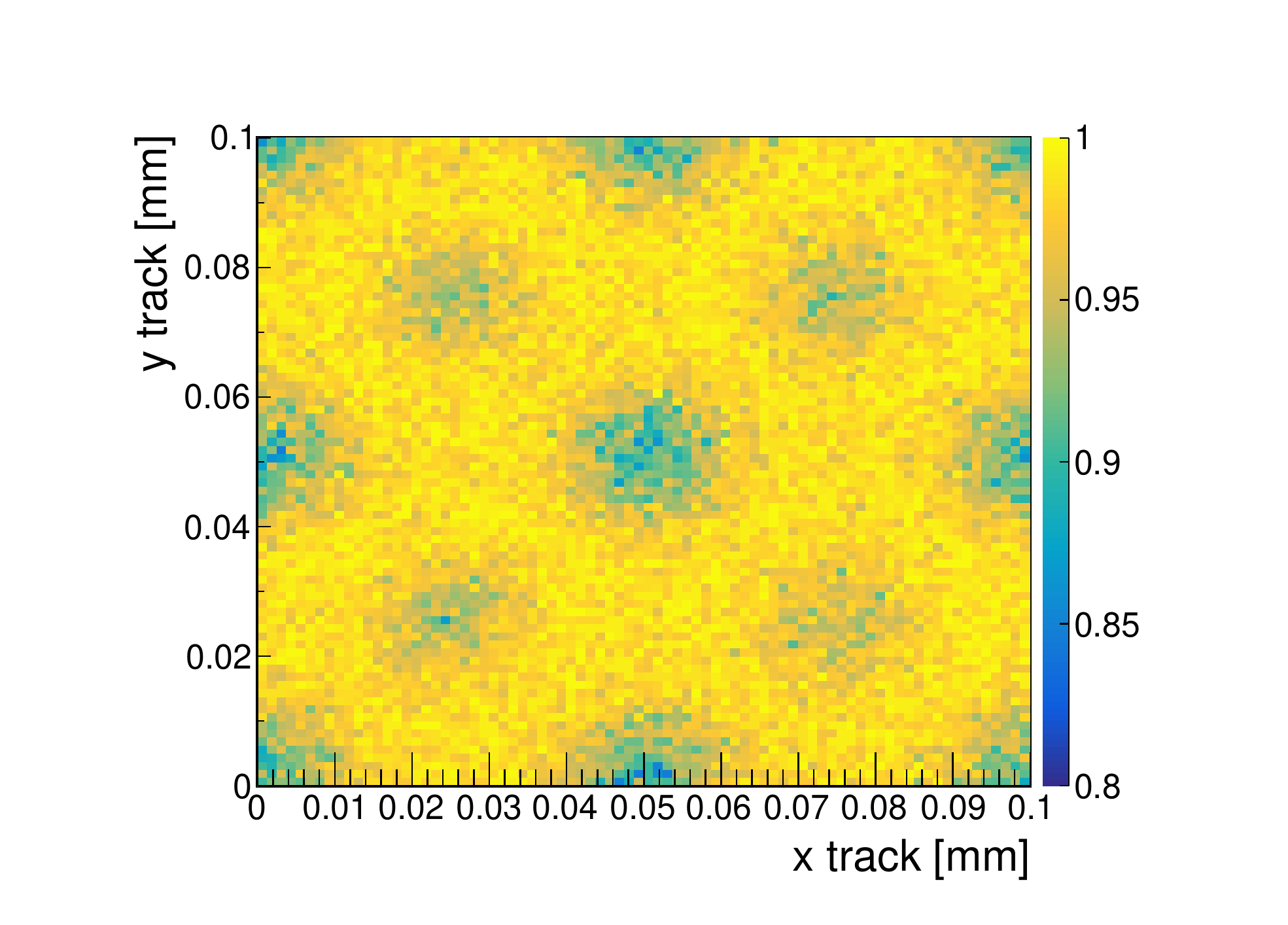}
    }
    \caption{Hit detection efficiencies before (a) and after irradiation (b) for a
    $50\times50\,\mu{\rm m}^2$ module.}
    \label{fig:s50x50}
\end{figure}

\begin{figure}[!htbp] 
    \centering
    {
        \includegraphics[width=0.5\linewidth,height=0.45\linewidth]{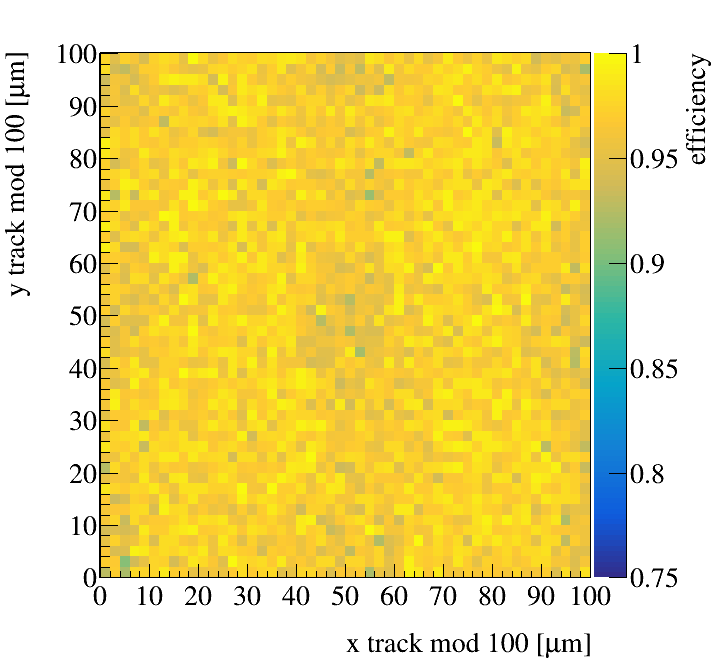}
    }
    \caption{Hit detection efficiencies when the sensor planes were tilted with respect to
    the incident particles by \SI{12}{\degree}. This figure corresponds to a \SI{50 x 50}{\micro\metre\square} 
    IMB-CNM sensor irradiated with a fluence of \SI{3e15}{\neq\centi\metre\tothe{-2}}, and bump-bonded to
    a ROC4SENS. The bias voltage was \SI{150}{\volt}.}
    \label{fig:tilted_sensors}
\end{figure}

After irradiation, a bias voltage of at least \SI{120}{\volt} was needed to reach 
similar efficiency. Figure~\ref{fig:s25x100} shows the comparison of hit 
efficiency before and after irradiation for \SI{25 x 100}{\micro\metre\square},
and figure~\ref{fig:s50x50} for \SI{50 x 50}{\micro\metre\square} pixel size modules,
and perpendicular incident tracks. In the efficiency plots the hits reconstructed 
over the whole module are projected on a  \num{2 x 2} pixel cell window to 
put in evidence the sensor geometry and the possible effects of the columnar electrodes. 
The geometrical inefficiency due to the columnar electrode diameter 
of \SI{5}{\micro\metre} was estimated to be  around \SI{1.5}{\percent}. 
\begin{table}[!tbp]
    \centering
    \caption{\label{tab:eff} Hit detection efficiency summary table for a fluence of \SI{1e16}{\neq\centi\metre\tothe{-2}} (errors are not quoted).}
    \smallskip
    \begin{tabular}{lcc}
        \hline
        3D Pixel-RD53A Linear FE&\SI{25 x 100}{\micro\metre\square}& \SI{50 x 50}{\micro\metre\square}\\
        \hline
        \hline
        Before irradiation & \SI{97.3}{\percent} & \SI{98.6}{\percent}\\
        After irradiation & \SI{96.6}{\percent} & \SI{97.5}{\percent}\\
        \hline
    \end{tabular}
\end{table}
This effect can be greatly reduced by tilting the module on the beam.  
Hit efficiency higher than \SI{99.3}{\percent} was recovered when sensors are tilted with respect
to the incident particles in order to maximize the charge sharing between pixels, as shown
in figure~\ref{fig:tilted_sensors}. This proves that the tested 3D sensors, both \SI{25 x 100}{\micro\meter\square} 
and \SI{50 x 50}{\micro\metre\square} are radiation tolerant up to the extreme expected doses at the HL-LHC. 
In particular, in the case of the \SI{25 x 100}{\micro\meter\square}, 1 electrode is enough to reach 
high hit efficiency. ROC4SENS modules were tested at the DESY test beam facility~\cite{desy} with 
\SI[per-mode=symbol]{5}{\giga\electronvolt\per\C} electrons. 
As expected no significant reduction of the sensor hit efficiency was observed after 
irradiation and overall efficiency of \SI{98}{\percent} at normal incidence was reached.

The hit detection efficiencies as calculated in our data analysis for different 
runs are reported in table~\ref{tab:eff}. 

\begin{figure}[!htbp] 
    \centering
    {
        \includegraphics[width=0.75\linewidth]{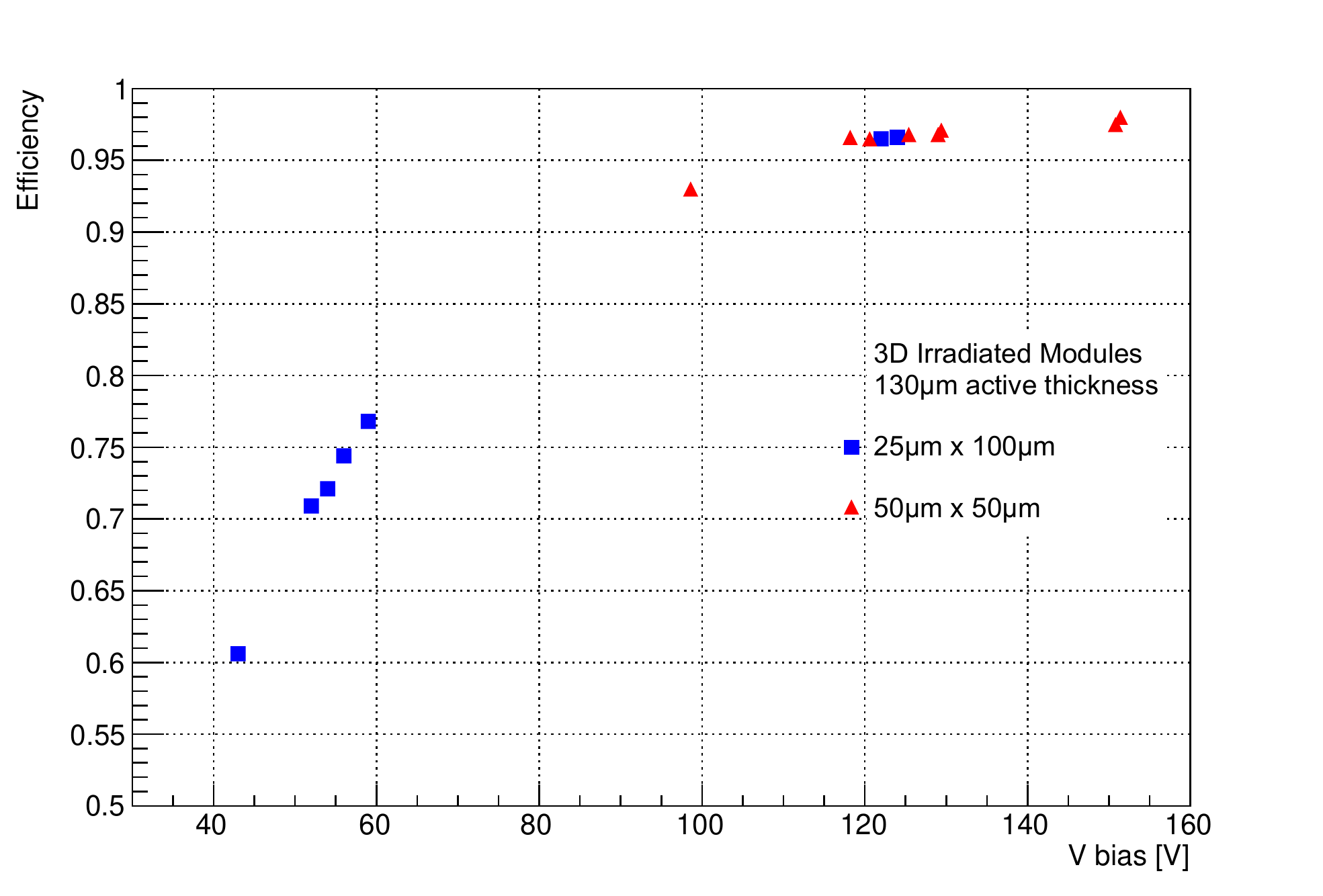}
    }
    \caption{Hit detection efficiency as a function of the bias voltage for irradiated sensors up to a fluence of 
    \SI{1e16}{\neq\centi\metre\square}; for \SI{50 x 50}{\micro\metre\square} (red triangles)
    and  \SI{25 x 100}{\micro\metre\square} 1E (blue squares), with perpendicular
    incident particles.}
    \label{fig:eff_vs_v}
\end{figure}

The hit efficiency as a function of the applied bias voltage is reported in 
figure~\ref{fig:eff_vs_v}, where it can be observed that starting from \SI{120}{\volt} 
the sensors reach the full depletion regime.

\section{Conclusions and Outlook}
Initial test beam results obtained with 3D pixel sensors show no significant 
degradation after proton irradiation up to \SI{1e16}{\neq\centi\metre\square} 
at bias voltages below \SI{200}{\volt}, confirming 3D pixel sensors as a possible robust 
option for the inner layers of future tracking detectors. In particular, it is shown that
one single collecting electrode per cell is enough to assure high efficiency.
New data is currently being analyzed from the CERN 2018 test beam campaign as well as from Fermilab 
and DESY test beams, in order to confirm and extend this study. At the beginning of 2019, a new 3D 
sensor batch was in production at FBK and a new batch with \SI{25 x 100}{\micro\metre\square} and
\SI{50 x 50}{\micro\metre\square} pitch sensors just finished at IMB-CNM. These will
be tested and characterized throughout 2019.

\section*{Acknowledgments}
I wish to thank Nuria Castello-Mor for providing me the elegant template
I used for the poster. We thank the RD53 Collaboration for the RD53A 
chip; we remind our results are not on chip performance but on sensor performance. 
The RD50 Collaboration for its support. Bonn ATLAS group for SCC cards and 
support for flip-chipping. 
Some of the measurements leading to these results have been performed at the CERN North Area
Test Beam Facility at Pr\'evessin (France).
Some of the measurements leading to these results have been performed at the Test Beam 
Facility at DESY Hamburg (Germany), a member of the Helmholtz Association (HGF).
This project  has been partially supported by the Spanish Ministry of Science under grants FPA2015-71292-C2-2-P,
FPA2017-85155-C4-1-R and FPA2017-85155-C4-2-R; and the European Union's Horizon 2020 Research 
and Innovation programme under Grant Agreement no. 654168 (AIDA-2020).
\bibliography{vci2019_899}
\end{document}